\documentclass[twocolumn,prc,showpacs,preprintnumbers,superscriptaddress,amsmath,amssymb]{revtex4}

\usepackage{graphicx}
\usepackage{dcolumn}
\usepackage{bm}

\usepackage{color}

\begin{document}


\preprint{1}

\title{Collective flows of $\alpha$-clustering $^{12}\mathrm{C}$ + $^{197}\mathrm{Au}$ by using different flow analysis methods}

\author{S. Zhang}
\affiliation{Shanghai Institute of Applied Physics, Chinese Academy of Sciences, Shanghai 201800, China}
\author{Y. G. Ma\footnote{Author to whom all correspondence should be addressed: ygma@sinap.ac.cn}}
\affiliation{Shanghai Institute of Applied Physics, Chinese Academy of Sciences, Shanghai 201800, China}
\affiliation{University of Chinese Academy of Sciences, Beijing 100049, China}
\affiliation{ShanghaiTech University, Shanghai 200031, China}
\author{J. H. Chen}
\affiliation{Shanghai Institute of Applied Physics, Chinese Academy of Sciences, Shanghai 201800, China}
\author{W. B. He}
\affiliation{Institute of Modern Physics, Fudan University, Shanghai 200433, China}
\author{C. Zhong}
\affiliation{Shanghai Institute of Applied Physics, Chinese Academy of Sciences, Shanghai 201800, China}

\date{\today}

\begin{abstract}
Recently the ratio of triangular flow to the elliptic flow ($v_3/v_2$) of hadrons was proposed as a probe to detect the pattern of $\alpha$-clustering $^{12}\mathrm{C}$ in $^{12}\mathrm{C}$+$^{197}\mathrm{Au}$ collisions at relativistic energy by a participant plane method (Phys. Rev. C 95, 064904 (2017)). In experimental event plane method, Q-cumulant method and two-particle correlation method with rapidity gap always were used for measurement of collective flow only by means of momentum space. By comparing collective flow through the different methods, the ratio of $v_3/v_2$ could be taken as an experimental probe to distinguish different $\alpha$-clustering structure of $^{12}\mathrm{C}$.
\end{abstract}

\maketitle

\section{Introduction}

In relativistic heavy-ion collisions, initial geometry distribution can affect some observables, such as collective flows~\cite{STARv1BES,STARv2BES, STARv3,STARFlowCME}, Hanbury-Brown-Twiss (HBT) correlation~\cite{STAR-HBTBES,PHENIX-HBTBES} and fluctuation~\cite{STAR-FLUCBES}. The initial geometry distribution was always influenced by initial dynamical fluctuation and intrinsic structure in the collided nuclei. These effects have been extensively investigated by different models, such as  hydrodynamical models~\cite{InitGeoHydo-1,InitGeoHydo-2, InitGeoHydo-3, InitGeoHydo-4} as well as in transport models~\cite{AlphaClusterHIC-1,AlphaClusterHIC-2,AMPTInitFluc-1,AMPTInitFluc-2,AMPTInitFluc-3},  respectively. Initial fluctuation effects have been  also proposed on some observables or physics quantities, such as on collective flows~\cite{NSTSongFlow,NSTWangFlow,GuoCC}, conserved quantities~\cite{NSTLuoCQ}, density fluctuations~\cite{NSTKeDensity}, and charge seperation \cite{ShouQY}.
In References~\cite{AlphaClusterHIC-1,AlphaClusterHIC-2} the carbon was considered with 3-$\alpha$ structure and collided against a heavy nucleus at very high energies and the results implied the final collective flow was sensitive to the initial geometry distribution. 

In one of our recent papers~\cite{SZhang-alphaC} collective flow ratio of $v_3/v_2$ (here $v_3$ is the triangular flow and $v_2$ is elliptic flow) was proposed as a probe to detect the intrinsic structure of $\alpha$-clustering nuclei with an $\alpha$-clustered $^{12}\mathrm{C}$ colliding against heavy-ion by using a multi-phase transport (AMPT) model. The $\alpha$-cluster  model which was originally proposed by Gamow~\cite{GamowAlphaModel} considered some light nuclei made of N-$\alpha$, such as $^{12}\mathrm{C}$ with 3-$\alpha$ and $^{16}\mathrm{O}$ with 4-$\alpha$. It was suggested that $\alpha$ clustering configurations can be identified by giant dipole resonance ~\cite{AlphaModelHe1,AlphaModelHe2} or photonuclear reaction in quasi-deuteron region ~\cite{AlphaModelHuang1,AlphaModelHuang2}  by EQMD model calculations.  Theoretically,   $^{12}\mathrm{C}$ could exhibit triangular or chain distribution of three $\alpha$ clusters and $^{16}\mathrm{O}$ could present kite, chain or square arrangement of four $\alpha$ clusters in some specific conditions. 

In our previous work~\cite{SZhang-alphaC} the participant plane (PP-) method~\cite{PartPlane-1,PartPlane-2,PartPlane-3} was employed to calculate the collective flow. The PP-method always was used in theoretical analysis for initial geometry fluctuation effect on collective flow. Event plane (EP-) method~\cite{Collectivity1992,flowMethod1998,EP-Method-1,EP-QC-Method-1,EP-Method-2}, Q-cumulant (QC-) method~\cite{EP-QC-Method-1,QC-method-1,AMPTInitFluc-1,AMPTInitFluc-2} and two particle correlation (2PC-) method with rapidity gap~\cite{TwoPartCorrRap-1,TwoPartCorrRap-2,TwoPartCorrRap-3,TwoPartCorrRap-4,TwoPartCorrRap-5} were performed in this work to calculate collective flow of final charged hadrons as that in experiments. The results implied that centrality dependence of ratio of $v_3/v_2$ from 2PC-method was consistent with that from PP-method.

\section{ model and calculation methods}

The phase space of the collision system was simulated by a multi-phase transport model (AMPT)~\cite{AMPT2005}. AMPT was developed to simulate heavy-ion collisions in a wide colliding energy range from SPS to LHC and successful to describe physics in relativistic heavy-ion collision for RHIC~\cite{AMPT2005} and LHC~\cite{AMPTGLM2016}, including pion-HBT correlations~\cite{AMPTHBT}, di-hadron azimuthal correlations~\cite{AMPTDiH}, collective flow~\cite{STARFlowAMPT,AMPTFlowLHC} and strangeness production~\cite{NSTJinS,SciChinaJinS}. In this model, the initial state was simulated by HIJING model~\cite{HIJING-1,HIJING-2}, and the melted partons would interact with each other in a parton cascade model (ZPC)~\cite{ZPCModel}, and then hadrons formed by a simple quark coalescence model participate in hadronic rescattering through a relativistic transport (ART) model~\cite{ARTModel}. The initial nucleon distribution in $^{12}\mathrm{C}$ was configured in HIJING model~\cite{HIJING-1,HIJING-2}  originally with pattern of Woods-Saxon distribution. And the other two cases of configuration of $^{12}\mathrm{C}$ structure were performed  as, three $\alpha$ clusters either in chain structure or in triangle structure. The parameters for the $\alpha$-clustered $^{12}\mathrm{C}$ were from calculation by EQMD model~\cite{AlphaModelHe1,AlphaModelHe2,EQMD1996} and discussed in our previous work~\cite{SZhang-alphaC} in detail. Then we can obtaine the phase space in $^{12}\mathrm{C}$+$^{197}\mathrm{Au}$ collisions at $\sqrt{s_{NN}}$ = 200 GeV for flow calculation by using AMPT model.


The final particle azimuthal distribution can be expanded as~\cite{Collectivity1992,flowMethod1998},
\begin{eqnarray}
\begin{split}
E\frac{d^3N}{d^3p} = &\frac{1}{2\pi}\frac{d^2N}{p_Tdp_Tdy}\\
&\times\left(1+\sum_{i=1}^{N}2v_n\cos[n(\phi-\Psi_{RP})]\right),
\end{split}
\label{FlowExpansion}
\end{eqnarray}
where $E$ is the energy, $p_T$ is transverse momentum, $y$ is rapidity, $\phi$ is  azimuthal angle of the particle. $\Psi_{RP}$ is reaction plane angle. And the Fourier coefficients $v_n (n=1,2,3,...)$ are collective flow to characterize different orders of azimuthal anisotropies with the form,
\begin{eqnarray}
v_n=\left<\cos(n[\phi-\Psi_{RP}])\right>,
\label{FlowDef}
\end{eqnarray}
where the bracket $\left<\right>$ denotes statistical averaging over particles and events. The true reaction plane angle $\Psi_{RP}$ always is estimated by event plane angle~\cite{Collectivity1992,flowMethod1998,EP-Method-1,EP-QC-Method-1,EP-Method-2} or by participant plane angle~\cite{PartPlane-1,PartPlane-2,PartPlane-3}. The harmonic flow can be calculated with respect to participant plane angle or event plane angle, called participant plane (PP-) method and event plane (EP-) method, respectively. Some method avoiding to reconstruct the reaction plane were developed, such as Q-cumulant (QC-) method~\cite{EP-QC-Method-1,QC-method-1,AMPTInitFluc-1,AMPTInitFluc-2} and two particle correlation (2PC-) method with rapidity gap~\cite{TwoPartCorrRap-1,TwoPartCorrRap-2,TwoPartCorrRap-3,TwoPartCorrRap-4,TwoPartCorrRap-5}. 

In participant coordinates system, the participant plane angle $\Psi_n\{PP\}$ can be defined by the following equation~\cite{PartPlane-1,PartPlane-2,PartPlane-3},
\begin{eqnarray}
\Psi_n\{PP\}=\frac{\tan^{-1}\left(\frac{\left<r^2\sin\left(n\phi_{part}\right)\right>}{\left<r^2\cos\left(n\phi_{part}\right)\right>}\right)+\pi}{n},
\label{PartPlanDef}
\end{eqnarray}
where, $\Psi_n\{PP\}$ is the nth-order participant plane angle, $r$ and $\phi_{part}$ are coordinate position and azimuthal angle of participants in the collision zone at initial state, and the average $\left<\cdots\right>$ denotes density weighting. 
And then the harmonic flow coefficients with respect to participant plane angle are defined as,
\begin{eqnarray}
v_n\equiv\left<\cos(n[\phi-\Psi_n\{PP\}])\right>.
\label{FlowPPDef}
\end{eqnarray}

PP-method has been used to calculate collective flow in some theoretical works~\cite{InitGeoHydo-1,InitGeoHydo-2,InitGeoHydo-3,InitGeoHydo-4,AMPTInitFluc-1,AMPTInitFluc-2,AMPTInitFluc-3,PartPlane-1,PartPlane-2,PartPlane-3}.  And it always was applied to discuss the initial geometry fluctuation effect on collective flow since the participant plane angle $\Psi_n\{PP\}$ was constructed by initial energy distribution in coordinates space with the event-by-event fluctuation effects.

Event plane (EP-) method always was used in experimental analysis for harmonic flow coefficients and the event plane angle $\Psi_n\{EP\}$ was defined by~\cite{Collectivity1992,flowMethod1998,EP-Method-1,EP-QC-Method-1,EP-Method-2}
\begin{eqnarray}
\begin{split}
Q_x&\equiv\sum_i^{M}\omega_i\cos(n\phi_i),\\
Q_y&\equiv\sum_i^{M}\omega_i\sin(n\phi_i),\\
\Psi_n\{EP\}&=\frac{1}{n}\tan^{-1}\left(\frac{Q_y}{Q_x}\right),
\end{split}
\label{EPDef}
\end{eqnarray}
where $\phi_i$ and $\omega_i$ are azimuthal angle and weight for the $i$th particle, respectively. In this work, the weight was chosen as unit. The sums extended over all particles used in the event plane reconstruction. Then harmonic flow coefficients could be calculated by
\begin{eqnarray}
\begin{split}
v_n &= \frac{v_n^{obs}}{\mathrm{Res}\{\Psi_n\{EP\}\}},\\
v_n^{obs} &= \langle\cos(km(\phi-\Psi_n\{EP\}))\rangle,\\
\mathrm{Res}\{\Psi_n\{EP\}\} &= \langle\cos(km(\Psi_n\{EP\}-\Psi_{RP}))\rangle.
\end{split}
\label{FlowEPDef}
\end{eqnarray}
The angular brackets indicate an average over all particles in all events and $km=n$ in this work. The resolution of event plane angle $\mathrm{Res}\{\Psi_n\{EP\}\}$ owing to finite number of particles can be calculated by,
\begin{eqnarray}
\begin{split}
\mathrm{Res}\{\Psi_n\{EP\}\} &= \langle\cos(km(\Psi_n\{EP\}-\Psi_{RP}))\rangle\\
&=\frac{\sqrt{\pi}}{2\sqrt{2}}\chi_m\exp\left(-\chi_m^4/4\right)I_{km}\\
I_{km}&=I_{(k-1)/2}\chi_m^2/4+I_{(k+1)/2}\chi_m^2/4,
\end{split}
\label{ResEPDef}
\end{eqnarray}
$\chi_m$ could be estimated by sub-event method. The event used to calculate event pane angle would randomly be splited into two sub-events, event $A$ and $B$, with maximum difference of particle number equal to 1. $\chi_m$ from sub-event resolution $\langle\cos(km(\Psi_m^{A}-\Psi_m^{B}))\rangle$ multiplying $\sqrt{2}$ would be the $\chi_m$ for full event resolution $\mathrm{Res}\{\Psi_n\{EP\}\}$. The detail for this analysis can be found in references~\cite{flowMethod1998,EP-Method-1,EP-QC-Method-1,EP-Method-2}.

To avoid reconstructing the event plane and eliminating non-flow contribution, multi-particle correlation method was developed to calculate the harmonic flow, such as Q-cumulant (QC-) method~\cite{EP-QC-Method-1,QC-method-1,AMPTInitFluc-1,AMPTInitFluc-2} and  two particle correlation (2PC-) method with rapidity gap~\cite{TwoPartCorrRap-1,TwoPartCorrRap-2,TwoPartCorrRap-3,TwoPartCorrRap-4,TwoPartCorrRap-5}. The multi-particle cumulant can be calculated directly from a $Q$ vector,
\begin{eqnarray}
Q_n = \sum_{i=1}^{M}e^{in\phi_i},
\label{QVectDef}
\end{eqnarray}
where $\phi_i$ is azimuthal angle of particles in momentum space. The two- and four- particle cumulants can be written as
\begin{eqnarray}
\begin{split}
\langle2\rangle &= \langle e^{in(\phi_1-\phi_2)}\rangle = \frac{|Q_n|^2-M}{M(M-1)},\\
\langle4\rangle &= \langle e^{in(\phi_1+\phi_2-\phi_3-\phi_4)}\rangle\\
&=\{|Q_n|^4+|Q_{2n}|^2-2\mathrm{Re}[Q_{2n}Q_n^{*}Q_n^{*}]\\
&-2[2(M-2)|Q_{2n}|^2-M(M-3)]\}\\
&\quad /[M(M-1)(M-2)(M-3)].
\end{split}
\label{QCDef}
\end{eqnarray}
Then, the average over all events can be formulated as
\begin{eqnarray}
\begin{split}
\langle\langle2\rangle\rangle &= \langle\langle e^{in(\phi_1-\phi_2)}\rangle\rangle = \frac{\sum_{event}(W_{\langle2\rangle})_i\langle2\rangle_i}{\sum_{event}(W_{\langle2\rangle})_i},\\
\langle\langle4\rangle\rangle &= \langle\langle e^{in(\phi_1+\phi_2-\phi_3-\phi_4)}\rangle\rangle = \frac{\sum_{event}(W_{\langle4\rangle})_i\langle4\rangle_i}{\sum_{event}(W_{\langle4\rangle})_i}.
\end{split}
\label{QCEventDef}
\end{eqnarray}
And then the two- and four-particle cumulants, and the flow can be written as,
\begin{eqnarray}
\begin{split}
c_n\{2\} &= \langle\langle2\rangle\rangle,\\
c_n\{4\} &= \langle\langle4\rangle\rangle-2\times\langle\langle2\rangle\rangle^2,\\
v_n\{2\} & = \sqrt{c_n\{2\}},\\
v_n\{4\} & = \sqrt[4]{-c_n\{4\}}.
\end{split}
\label{CnFlowDef}
\end{eqnarray}
The above flow $v_n\{2\}$ and $v_n\{4\}$ is the reference flow with integration over transverse momentum $p_T$. The differential flow as a function of $p_T$ will not be discussed here. The detail for the algorithm of Q-cumulant method can be found in references~\cite{EP-QC-Method-1,QC-method-1,AMPTInitFluc-1,AMPTInitFluc-2}.

In the 2PC-method~\cite{TwoPartCorrRap-1,TwoPartCorrRap-2,TwoPartCorrRap-3,TwoPartCorrRap-4,TwoPartCorrRap-5}, the two-dimensional (2D) two-particle correlation function is generally defined as
\begin{equation}
C(\Delta\phi,\Delta\eta) = \frac{S(\Delta\phi,\Delta\eta)}{B(\Delta\phi,\Delta\eta)},
\end{equation}
where 
\begin{eqnarray}
\begin{split}
S(\Delta\phi,\Delta\eta) &= \frac{dN}{d\Delta\phi d\Delta\eta},\\
B(\Delta\phi,\Delta\eta) &= \frac{dN}{d\Delta\phi d\Delta\eta}.
\end{split}
\label{TwoPartCorrFuncDef}
\end{eqnarray}
$S(\Delta\phi,\Delta\eta)$ and $B(\Delta\phi,\Delta\eta)$ are the same-event pair distribution and the combinatorial distribution in two-particle phase space $(\Delta\phi=\phi_a-\phi_b,\Delta\eta=\eta_a-\eta_b)$, respectively. Mix-event method is employed to calculate $B(\Delta\phi,\Delta\eta)$. In mix-event the pair particles are from two different events with similar event properties such as number of track. To reduce non-flow contribution at $(\Delta\phi,\Delta\eta)\sim(0,0)$, a one-dimensional (1D) $\Delta\phi$ correlation function can be given with $2<|\Delta\eta|<5$.
The 1D two-particle correlation function is generally defined as
\begin{eqnarray}
C(\Delta\phi) &= A\frac{\int S(\Delta\phi,\Delta\eta)d\Delta\eta}{\int B(\Delta\phi,\Delta\eta)d\Delta\eta}.
\label{TwoPartCorrFuncDphiDef}
\end{eqnarray}
The normalization constant A is determined by scaling the number of pairs in $2<|\Delta\eta|<5$ to be the same between the same event counts ($S$) and the mix-event counts (B)~\cite{TwoPartCorrRap-1}. And the distribution of pairs in $\Delta\phi$ can be expanded into a Fourier series,
\begin{eqnarray}
\frac{dN_{pairs}}{d\Delta\phi}\propto 1+2\sum_{n=1}^{\infty}v_{n,n}\left(p_T^a,p_T^b\right)\cos (n\Delta\phi).
\label{dNPairExp}
\end{eqnarray}
The coefficients $v_{n,n}$ can be calculated directly by
\begin{eqnarray}
v_{n,n} = \langle\cos (n\Delta\phi)\rangle = \frac{\sum_{m=1}^N\cos(n\Delta\phi_m)C(\Delta\phi_m)}{\sum_{m=1}^NC(\Delta\phi)},
\label{vnnDef}
\end{eqnarray}
where $n=2,3$, and N=200 is the number of $\Delta\phi$ bins. The harmonic flow coefficients $v_n$ ($n=2,3$) can be calculated as $v_n = v_{n,n}/\sqrt{|v_{n,n}|}$. The detail of this method can be found from references~\cite{TwoPartCorrRap-1}.

The above introduced method will be applied to calculate the harmonic flow in $^{12}\mathrm{C}$+$^{197}\mathrm{Au}$ collisions at $\sqrt{s_{NN}}$ = 200 GeV for different configuration of $^{12}\mathrm{C}$ structure.

\section{Results and discussion}


\begin{figure}[htbp]
\includegraphics[width=8.6cm]{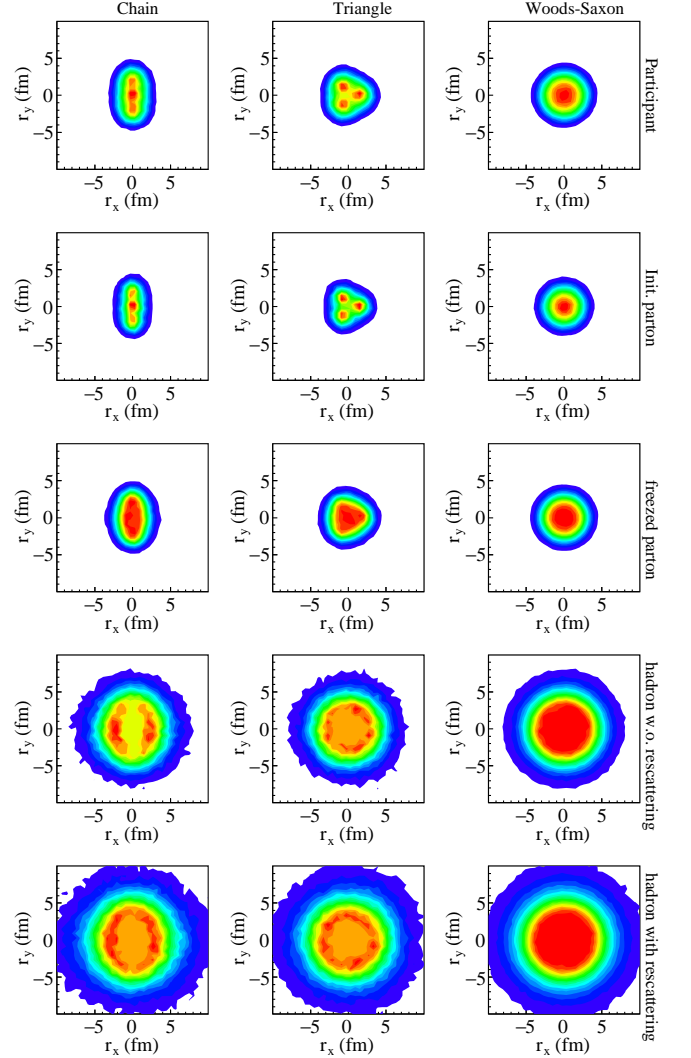}
\caption{
\label{fig:xyDstEvolution}(color online)
 In $^{12}\mathrm{C}$+$^{197}\mathrm{Au}$ collisions at $\sqrt{s_{NN}}$ = 200 GeV, the distribution in X-Y plane of participants (first row), initial partons (second row), freezed partons (third row), hadrons without hadronic rescattering (fourth row) and hadrons with hadronic rescattering (fifth row) for different $\alpha$-clustering configuration of $^{12}\mathrm{C}$ structure, i.e., chain (left column), triangle (middle column),  and Woods-Saxon (right column) structures.
 }
\end{figure}

Figure~\ref{fig:xyDstEvolution} shows the distribution in X-Y plane for participants, initial partons, freezed partons, hadrons without hadronic rescattering and hadron with hadronic rescattering in $^{12}\mathrm{C}$+$^{197}\mathrm{Au}$ collisions at $\sqrt{s_{NN}}$ = 200 GeV for different configuration of $^{12}\mathrm{C}$ structure. The initial nucleon distribution in $^{12}\mathrm{C}$ is configured for (a) three $\alpha$ clusters in Chain structure, (b) three $\alpha$ clusters in Triangle structure, and (c) nucleons in Woods-Saxon distribution from HIJING model~\cite{HIJING-1,HIJING-2} (Woods-Saxon). The distribution of radial centre of the $\alpha$ clusters in $^{12}\mathrm{C}$ is assumed to be a Gaussian function, $e^{-0.5\left(\frac{r-r_c}{\sigma_{r_c}}\right)^{2}}$, here $r_c$ is average radial center of an $\alpha$ cluster and $\sigma_{r_c}$ is the width of the distribution. And the nucleon inside each  $\alpha$ cluster will be given by Woods-Saxon distribution. The parameters of $r_c$ and $\sigma_{r_c}$ can be obtained from the EQMD calculation~\cite{AlphaModelHe1,AlphaModelHe2,AlphaModelHuang1,AlphaModelHuang2}.  For  the Triangle structure,   $r_c$ = 1.8 fm and $\sigma_{r_c}$  = 0.1 fm. For Chain structure, $r_c$ = 2.5 fm, $\sigma_{r_c}$  = 0.1 fm for two $\alpha$ clusters and the other one will be at the center in $^{12}\mathrm{C}$. Once the  radial centre of the $\alpha$ cluster is determined, the centers of the three clusters will be placed in an equilateral triangle for the Triangle structure or in a line for Chain structure.

\begin{figure}[htbp]
\includegraphics[width=8.5cm]{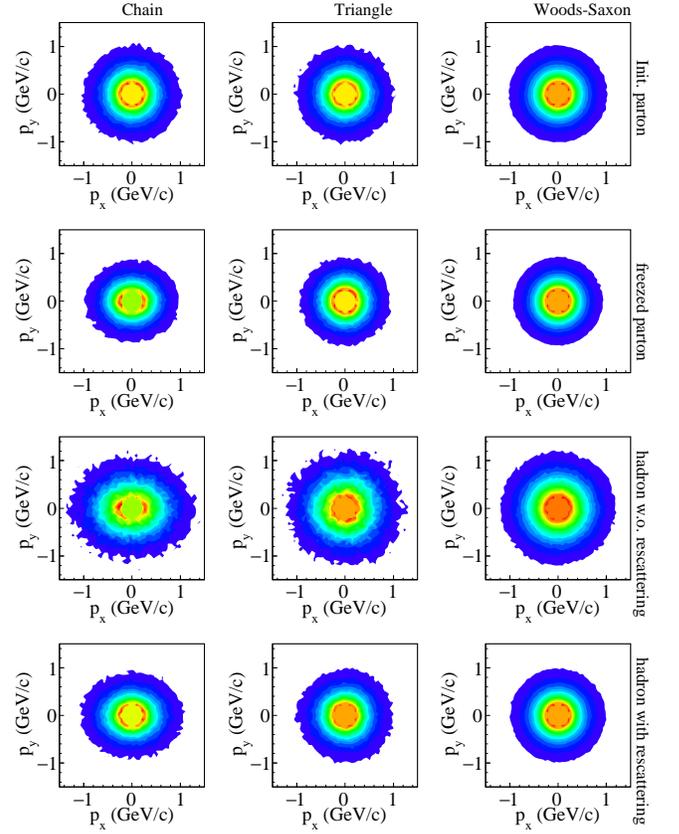}
\caption{
\label{fig:PxPyDstEvolution}(color online)
 In $^{12}\mathrm{C}$+$^{197}\mathrm{Au}$ collisions at $\sqrt{s_{NN}}$ = 200 GeV, the distribution in $p_x$-$p_y$ plane of initial partons (first row), freezed partons (second row), hadrons without hadronic rescattering (third row) and hadrons with hadronic rescattering (fourth row) for different $\alpha$-clustering configuration of $^{12}\mathrm{C}$ structure, i.e., chain (left column), triangle (middle column),  and Woods-Saxon (right column) structures.
}
\end{figure}

The participant plane angle $\Psi_n\{PP\}$ is reconstructed by using the coordinates from initial parton (the second row panels) instead of participants (the first row panels) for considering dynamical fluctuation from the initial collisions in HIJING model. These plots show X-Y coordinates distribution with respective to participant plane angle $\Psi_n\{PP\}$ and the events are selected for impact parameter $b$ = 0 fm. After parton cascade (the third row panels), hadronization (the fourth row panels) and hadronic rescattering (the fifth row panels), the coordinates distributions in X-Y plane of production gradually tend to be thermalized. Similarly it was calculated for the distribution in $p_x$-$p_y$ plane as shown in figure~\ref{fig:PxPyDstEvolution} with transverse momentum $p_T$ window (0.2,3) GeV/$c$ and rapidity window (-1,1). Note that there is no $p_x$-$p_y$ distribution of participants since their momenta along beam direction (z-axis). For the chain structure, we can see that the $p_x$-$p_y$ distribution is symmetrical for the initial partons and becomes asymmetrical after parton cascade (freezed parton) and this asymmetrical distribution is inherited by hadrons (without and with hadronic rescattering). For the triangle and Woods-Saxon distribution, there is no obvious evolution of azimuthal asymmetry at different stage which will be investigated by harmonic flow coefficients in the following. The flow formation mechanism was investigated in some works~\cite{AMPTFlowEscape1,AMPTFlowEscape2}. The escape mechanism and contribution form parton collision time to collective flow were suggested in these work. From these viewpoints, $p_x$-$p_y$ distribution will evolve from geometrical asymmetry to momentum asymmetry in transverse plane, which is obvious in the chain structure or in semi-central Au+Au collisions. And the following results will be calculated for final charged hadrons to investigate harmonic flow coefficients as did in experiments.

\begin{figure}[htbp]
\includegraphics[width=8.5cm]{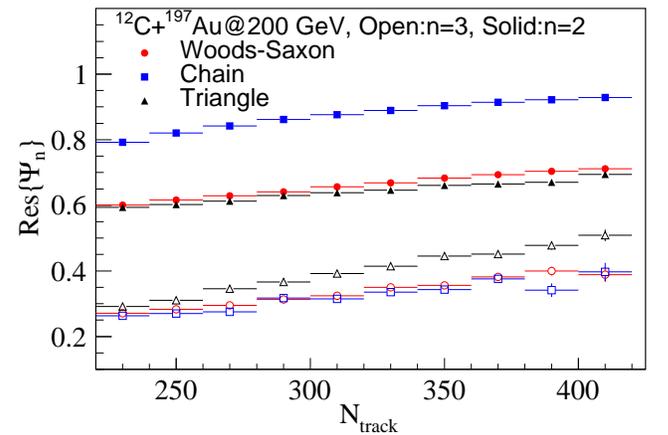}
\caption{
\label{fig:ResEP}(color online)
 In $^{12}\mathrm{C}$+$^{197}\mathrm{Au}$ collisions at $\sqrt{s_{NN}}$ = 200 GeV, the second and third order event plane angle resolution as a function of $\mathrm{N_{track}}$ for different configuration of $^{12}\mathrm{C}$ structure.
 }
\end{figure}

Figure~\ref{fig:ResEP} presents the second and third order event plane angle resolution $\mathrm{Res}\{\Psi_n\{EP\}\}$ as a function of $\mathrm{N_{track}}$ in $^{12}\mathrm{C}$+$^{197}\mathrm{Au}$ collisions at $\sqrt{s_{NN}}$ = 200 GeV for different configuration of $^{12}\mathrm{C}$ structure.
$\mathrm{N_{track}}$ is calculated in the rapidity window ($-2<y<2$) and transverse momentum window ($0.2<p_T<6$) GeV/$c$ for the charged pion ($\pi^{\pm}$), Kaon ($K^{\pm}$) and proton ($p$, $\bar{p}$). The event plane angle resolution $\mathrm{Res}\{\Psi_n\{EP\}\}$ increases with the increasing of $\mathrm{N_{track}}$. The second (third) order event plane resolution is higher for chain (triangle) structure than for other patterns. These indicate that the event plane angle resolution depends not only on the number of particles used to reconstruct the event plane but also on the harmonic flow coefficients for the event plane angle which is determined by the harmonic flow coefficients itself~\cite{flowMethod1998,EP-Method-1,EP-QC-Method-1,EP-Method-2}.

\begin{figure}[htbp]
\includegraphics[width=8.5cm]{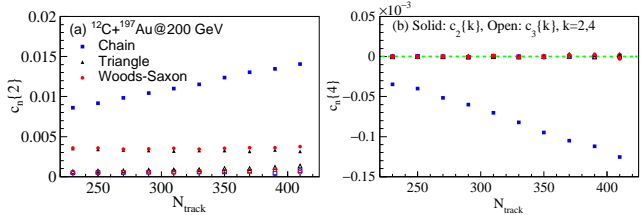}
\caption{
\label{fig:CnQC}(color online)
 In $^{12}\mathrm{C}$+$^{197}\mathrm{Au}$ collisions at $\sqrt{s_{NN}}$ = 200 GeV, the two- and four-particle cumulants $c_n\{2\}$ and $c_n\{4\}$ as a function of $\mathrm{N_{track}}$ for different configuration of $^{12}\mathrm{C}$ structure.
 }
\end{figure}

For the Q-cumulant (QC-) method, we investigated the two- and four-particle cumulants $c_n\{2\}$ and $c_n\{4\}$ as a function of $\mathrm{N_{track}}$ in $^{12}\mathrm{C}$+$^{197}\mathrm{Au}$ collisions at $\sqrt{s_{NN}}$ = 200 GeV for different configurations of $^{12}\mathrm{C}$ structure, as shown in figure~\ref{fig:CnQC}. From panel (a) in figure~\ref{fig:CnQC}  we can see that $c_2\{2\}$ increases rapidly for chain structure and $c_3\{2\}$ slightly increases with the increasing of $\mathrm{N_{track}}$ for triangle structure, respectively, and $c_n\{2\}$ ($n=2,3$) keeps almost flat for other cases. Unfortunately the four-particle cumulants $c_n\{4\}$ only give the reasonable value ($<0$ for $v_n\{4\}$ in formula~(\ref{CnFlowDef})) of $c_2\{4\}$ in chain structure pattern then only $v_n\{2\}$ is presented by QC-method. Obviously this method for calculation of multi-particle cumulants depends not only on the $\mathrm{N_{track}}$ but also on the asymmetrical flow itself~\cite{EP-QC-Method-1,QC-method-1,AMPTInitFluc-1,AMPTInitFluc-2}.

\begin{figure}[htbp]
\includegraphics[width=8.5cm]{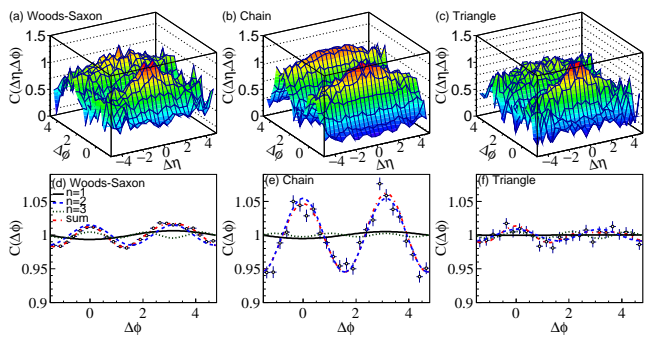}
\caption{
\label{fig:CDetaDphi}(color online)
 In $^{12}\mathrm{C}$+$^{197}\mathrm{Au}$ collisions at $\sqrt{s_{NN}}$ = 200 GeV, the two particle correlation function (rebinned into 25 bins) for different configuration of $^{12}\mathrm{C}$ structure, $\mathrm{N_{track}}>$200 and $1<p_T<3$ GeV/$c$.
 }
\end{figure}

Figure~\ref{fig:CDetaDphi} shows the 2D $\Delta\eta-\Delta\phi$ and 1D $\Delta\phi$ two particle correlation functions for different configurations of $^{12}\mathrm{C}$ structure with $\mathrm{N_{track}}>$200 and $1<p_T<3$ GeV/$c$ in $^{12}\mathrm{C}$+$^{197}\mathrm{Au}$ collisions at $\sqrt{s_{NN}}$ = 200 GeV. The short range correlation at $(\Delta\phi,\Delta\eta)\sim(0,0)$ suggests that there are autocorrelations from jet fragmentation and resonance decays~\cite{TwoPartCorrRap-1,TwoPartCorrRap-2,TwoPartCorrRap-3,TwoPartCorrRap-4,TwoPartCorrRap-5} as shown at panels (a), (b) and (c) in figure~\ref{fig:CDetaDphi}, so-called the non-flow contribution. We can see that the correlation function of Woods-Saxon (panel a) case is similar to that of Triangle (panel c) case. This similar distribution was found in $px-py$ distribution of final particles shown in figure~\ref{fig:PxPyDstEvolution} (bottom line). The panel (b) in figure~\ref{fig:CDetaDphi} presents strong correlation near $\Delta\phi$ =0 and $\Delta\phi$ = $\pi$, which is mainly from the flow contribution. The non-flow contribution was reflected near $(\Delta\phi,\Delta\eta)\sim(0,0)$ since jet quenching and resonance decay always happen in a narrow $(\Delta\phi,\Delta\eta)$ space, namely short range correlation.
To eliminate the non-flow contribution,1D $\Delta\phi$ correlation functions presented at panels (d), (e) and (f) in figure~\ref{fig:CDetaDphi} are obtained through integrating 2D $\Delta\eta-\Delta\phi$ correlation function with a large $\Delta\eta$ gap ($\Delta\eta>2$). At panels (d), (e) and (f), the lines are contributions from the individual $v_{n,n}$ components  ($n=1,2,3$) and their sum as in formula~(\ref{dNPairExp}) and the markers are from AMPT model. And then the harmonic flow can be calculated by using formula~(\ref{vnnDef}). 

\begin{figure}[htbp]
\includegraphics[width=8.5cm]{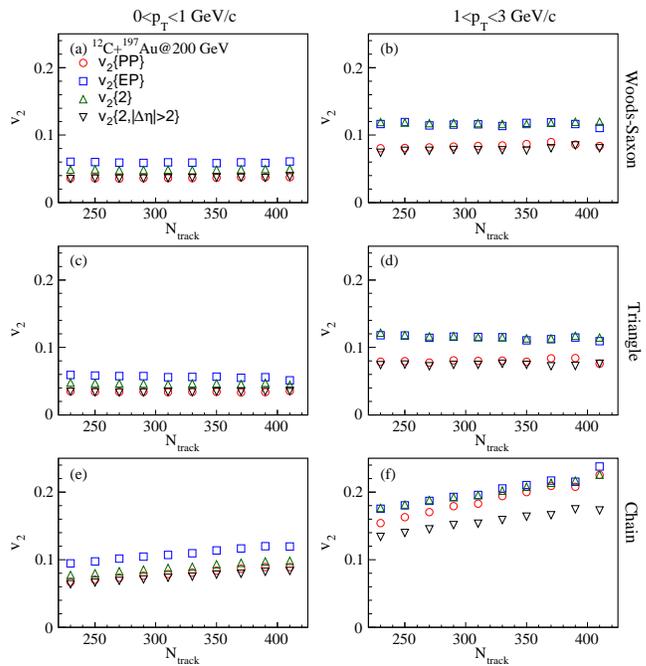}
\caption{
\label{fig:Flowv2}(color online)
 In $^{12}\mathrm{C}$+$^{197}\mathrm{Au}$ collisions at $\sqrt{s_{NN}}$ = 200 GeV, the elliptic flow $v_2$ by using different methods for different configuration of $^{12}\mathrm{C}$ structure.
 }
\end{figure}

\begin{figure}[htbp]
\includegraphics[width=8.5cm]{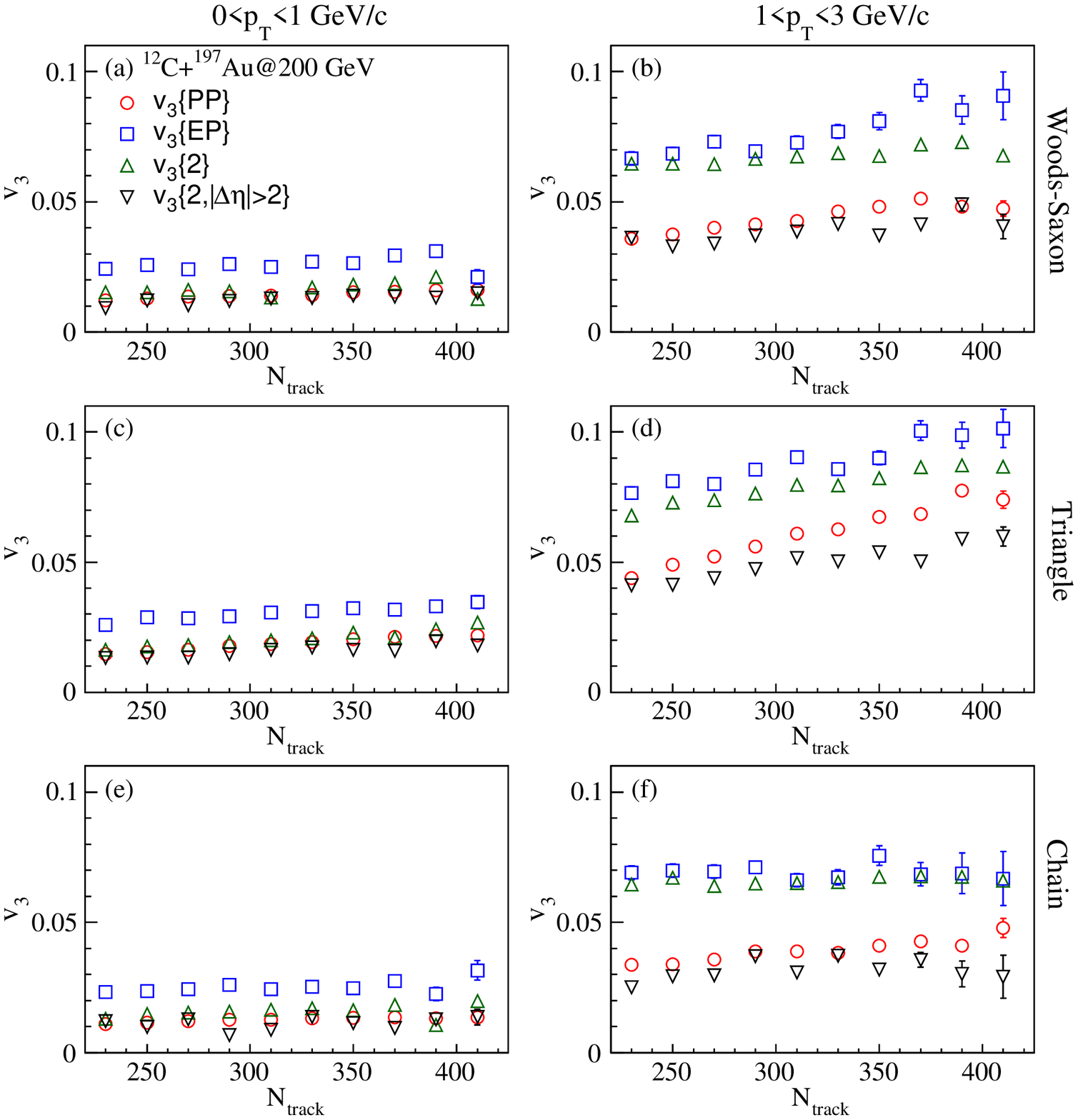}
\caption{
\label{fig:flowv3}(color online)
 In $^{12}\mathrm{C}$+$^{197}\mathrm{Au}$ collisions at $\sqrt{s_{NN}}$ = 200 GeV, the trianglular flow $v_3$ by using different methods for different configuration of $^{12}\mathrm{C}$ structure.
 }
\end{figure}

With the above introduced method for flow calculation, the elliptic and triangular flow are calculated and presented in figure~\ref{fig:Flowv2} and figure~\ref{fig:flowv3} separately. $v_n\{PP\}$, $v_n\{EP\}$, $v_n\{2\}$ and $v_n\{2,|\Delta\eta|>2\}$ denote the harmonic flow coefficients are calculated by PP-method, EP-method, QC-method and 2PC-method, respectively. Note that there is no $v_n\{4\}$ results by QC-method since the cumulant of $c_n\{4\}$ is unreasonable but in chain structure as shown in figure~\ref{fig:CnQC} (b). The trend of $\mathrm{N_{track}}$ dependence for the elliptic flow by using different method are consistent with each other. The elliptic flow (figure~\ref{fig:Flowv2}) in Woods-Saxon and triangle cases keep a flat $\mathrm{N_{track}}$ dependence in different $p_T$ ranges. The elliptic flow in chain case presents slightly increasing trend with the increasing of $\mathrm{N_{track}}$ in lower $p_T$ range ($0<p_T<1$ GeV/$c$) and the increasing trend becomes obvious in higher $p_T$ range  ($1<p_T<3$ GeV/$c$).
And $v_2$ is higher by EP-method than by other methods in lower $p_T$ range ($0<p_T<1$ GeV/$c$). In Woods-Saxon distribution and triangle pattern, in higher $p_T$ range  ($1<p_T<3$ GeV/$c$) $v_2\{EP\}$ is approximately equal to $v_2\{2\}$ and $v_2\{2,|\Delta\eta|>2\}$ to $v_2\{PP\}$. However, in chain pattern, $v_2\{2,|\Delta\eta|>2\}$ is lower than elliptic flow by using other methods in higher $p_T$ range. 
 Figure~\ref{fig:flowv3} shows the triangular flow as a function of $\mathrm{N_{track}}$ in different $p_T$ range by the above flow analysis methods. The triangular flow presents flat $\mathrm{N_{track}}$ dependent trend in lower $p_T$ range ($0<p_T<1$ GeV/$c$) and also in higher $p_T$ range ($1<p_T<3$ GeV/$c$) for chain structure. The triangular flow increases with the increasing of $\mathrm{N_{track}}$ in higher $p_T$ range ($1<p_T<3$ GeV/$c$) for Woods-Saxon distribution and triangle structure.
The elliptic and triangular flow increase with $\mathrm{N_{track}}$ in higher $p_T$ range ($1<p_T<3$ GeV/$c$) for some cases in ~\ref{fig:Flowv2} and figure~\ref{fig:flowv3}. This phenomenon is related to the transformation from the coordinate anisotropy to momentum anisotropy. The participant eccentricity coefficients is defined as $\epsilon_n\{PP\}\equiv\frac{\sqrt{\left<r^2\cos\left(n\phi_{part}\right)\right>^2+\left<r^2\sin\left(n\phi_{part}\right)\right>^2}}{\left<r^2\right>}$. The $\epsilon_2\{PP\}$ and $\epsilon_3\{PP\}$ were calculated in the impact parameter $b$ range (0,4) fm corresponding to $\mathrm{N_{track}}$ range (220,440) and there was no obvious $b$ ($\mathrm{N_{track}}$) dependence of $\epsilon_2\{PP\}$ and $\epsilon_3\{PP\}$. Table~\ref{tab:e2e3} shows the value of $\epsilon_2\{PP\}$ and $\epsilon_3\{PP\}$ for difference configuration cases. This implies that transformation from the coordinate anisotropy to momentum anisotropy depends on the number of particles created in the system, such as $\mathrm{N_{track}}$, which is consistent with that from reference~\cite{HydroKN-1,HydroKN-2,NSTSongFlow,AlphaClusterHIC-1}.
From different method comparing, it can be found that $v_3\{2,|\Delta\eta|>2\}$ takes the lowest value in triangle pattern in higher $p_T$ range for the non-flow contribution is omitted effectively.

\begin{table}
\caption{ \label{tab:e2e3} Participant eccentricity coefficients for different configuration cases. }
\begin{ruledtabular}
\begin{tabular}{lll}
 & $\epsilon_2\{PP\}$ & $\epsilon_3\{PP\}$\\
\hline
Chain & 0.5 & 0.16\\
Triangle & 0.23 & 0.25\\
Woods-Saxon & 0.26 & 0.22
\end{tabular}
\end{ruledtabular}
\end{table}

\begin{figure}[htbp]
\includegraphics[width=8.5cm]{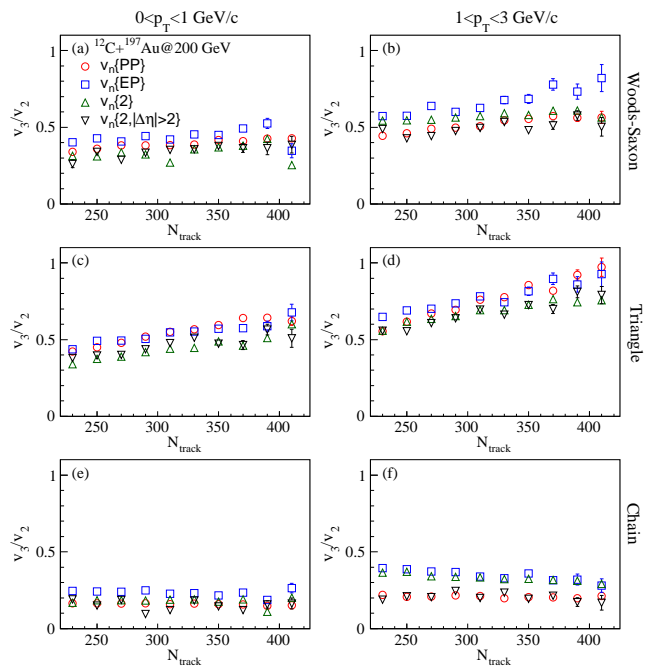}
\caption{
\label{fig:flowv3Overv2}(color online)
 In $^{12}\mathrm{C}$+$^{197}\mathrm{Au}$ collisions at $\sqrt{s_{NN}}$ = 200 GeV, the ratio of $v_3/v_2$ by using different methods for different $\alpha$-clustering configuration of $^{12}\mathrm{C}$ structure.
 }
\end{figure}

As we mentioned in our previous work~\cite{SZhang-alphaC}, $v_3/v_2$ can be a probe to distinguish the geometry pattern of $^{12}\mathrm{C}$ through collective flow measurements. In this work we calculated the probe by using different flow calculation/measurement method as shown in figure~\ref{fig:flowv3Overv2}. In Woods-Saxon distribution pattern, the ratio of $v_3/v_2$ keeps flat with the increasing of $\mathrm{N_{track}}$ by using PP-method, QC-method and 2PC-method, but by using EP-method $v_3/v_2$ increases with the increasing of $\mathrm{N_{track}}$. In triangle pattern, $v_3/v_2$ presents increasing trend with the increasing of $\mathrm{N_{track}}$ by using different flow calculation method. And in chain pattern, the ratios of $v_3/v_2$ through PP-method and 2PC-method keep flat with the increasing of  $\mathrm{N_{track}}$, but the ratios through EP-method and QC-method slightly decrease with the increasing of $\mathrm{N_{track}}$. So we can conclude that the ratio of $v_3/v_2$ from the 2PC-method is closest to that from the PP-method and we propose to investigate the $\alpha$-clustering nuclear structure by flow measurement in heavy-ion collisions. To distinguish $\alpha$-clustering nuclear structure by $v_3/v_2$ in experiment, non-exotic structure nucleus near $^{12}\mathrm{C}$ colliding against $^{197}\mathrm{Au}$  can be a reference collision system. If $v_3/v_2$ in $^{12}\mathrm{C} + ^{197}\mathrm{Au}$ collisions is close to that in the reference system, $^{12}\mathrm{C}$ will be in non-exotic nuclear structure. If $v_3/v_2$ in $^{12}\mathrm{C} + ^{197}\mathrm{Au}$ collisions is obviously smaller than that in the reference system, $^{12}\mathrm{C}$ should be in three $\alpha$-clusters chain structure. And if $v_3/v_2$ in $^{12}\mathrm{C} + ^{197}\mathrm{Au}$ collisions is significantly increasing with  $\mathrm{N_{track}}$, $^{12}\mathrm{C}$ can be seen as a triangle shape with three $\alpha$-clusters.

\section{Summary}
In summary,  participant plane, event plane, Q-cumulant and two particle correlation with $\Delta\eta$ gap methods were employed to calculate the elliptic and triangular flow coefficients in $^{12}\mathrm{C}$+$^{197}\mathrm{Au}$ collisions at $\sqrt{s_{NN}}$ = 200 GeV with $\alpha$-clustering $^{12}\mathrm{C}$ structure arranged in triangle, chain  and Woods-Saxon distributions, respectively. The ratio of $v_3/v_2$ was proposed as a probe to distinguish the pattern of $\alpha$-clustered $^{12}\mathrm{C}$ structure. By using two-particle correlation method with $\Delta\eta$ gap the ratio is closest to that by using participant plane method. And $v_3/v_2$ can be measured in relativistic nucleus-nucleus collisions by two-particle correlation method with $\Delta\eta$ gap and as well as 4- or more-particle cumulant method is alternative if it works well for this small system.

\vspace{0.5cm}

We are grateful for discussion with Dr. Aihong Tang from BNL. This work was supported in part by  the National Natural Science Foundation of China under contract Nos. 11421505, 11220101005, 11775288 and U1232206, the Major State Basic Research Development Program in China under Contract No. 2014CB845400, and the Key Research Program of Frontier Sciences of the CAS under Grant No. QYZDJ-SSW-SLH002, and National Key R\&D Program of China under Grant No. 2016YFE0100900.


\bibliography{NuclStruEff-Flow-MethodIntro3}

\end{document}